\documentstyle[floats,prl,aps,epsf,preprint]{revtex}

\begin{document}
\bibliographystyle{plain}
\title{Bulk versus Brane in the Absorption and Emission : $5$D Rotating
Black Hole Case}
\author{Eylee Jung\footnote{Email:eylee@kyungnam.ac.kr} 
and
D. K. Park\footnote{Email:dkpark@hep.kyungnam.ac.kr 
}}
\address{Department of Physics, Kyungnam University,
Masan, 631-701, Korea.}
\date{\today}
\maketitle

\begin{abstract}
The absorption and emission spectra for the minimally-coupled brane and bulk scalar 
fields are numerically 
computed when the spacetime is a $5d$ rotating black hole carrying the two 
different angular momentum parameters $a$ and $b$. The effect of the superradiant
scattering in the spectra is carefully examined. 
It is shown that the low-energy limit of the 
total absorption cross section always equal to the area of the non-spherically 
symmetric horizon, {\it i.e.} $4\pi (r_H^2 + a^2)$ for the brane scalar and 
$2\pi^2 (r_H^2 + a^2) (r_H^2 + b^2)/r_H$ for the bulk scalar where
$r_H$ is an horizon radius. The energy amplification 
for the bulk scalar is roughly order of $10^{-9} \%$ while that for the brane scalar
is order of unity. This indicates that the effect of the superradiance is 
negligible for the case of the bulk scalar. Thus the standard claim that 
{\it black holes radiate mainly on the brane} is not changed although the effect 
of the superradiance is taken into account. The physical implication of this fact 
is discussed in the context of TeV-scale gravity. 
\end{abstract}
\newpage
The recent brane-world scenarios which assume the
large\cite{ark98-1,anto98} or warped\cite{rs99-1,rs99-2} extra dimensions
generally allow the emergence of 
the TeV-scale gravity, which opens the possibility to make the tiny black holes 
factory in the future high-energy 
colliders\cite{gidd02-1,dimo01-1,eard02-1,stoj04,card05}. In this reason the 
absorption and emission problems for the higher-dimensional non-rotating 
black holes were extensively explored 
recently\cite{harris03-1,jung04-1,jung05-1}. It was found\cite{harris03-1} 
numerically that the emission on the brane is dominant compared to the 
emission off the brane in the Schwarzschild black hole background. 
This fact strongly supports the
main conclusion of Ref.\cite{emp00}. Adopting a different numerical technique
used in Ref.\cite{sanc78,jung04-2}, it was also found\cite{jung05-1} that the
higher-dimensional charged black holes also radiate mainly on the brane if the 
number of the extra dimensions is not too large.

For the higher-dimensional rotating black holes, however, the situation 
can be more complicated. For the non-rotating black holes the crucial 
factor which makes the Hawking evaporation on the brane to be dominant is a 
geometrical factor $r_H / L << 1$, where $r_H$ is an horizon radius and $L$ is
a size of the extra dimensions. In the rotating black holes, besides this 
geometrical factor, there is 
another important factor called superradiance, 
which means that the incident wave is amplified by the extraction of the
rotation energy of the black holes under the particular condition. The effect
of the superradiance in the $4d$ black holes was extensively studied long
ago\cite{zeldo71-1,press72,star73-1,star74-1}. The black hole bomb, {\it i.e.}
rotating black hole plus mirror system, was recently re-examined in 
detail\cite{cardo04} from the aspect of the black hole stability.

The importance of the superradiance modes in the tiny rotating black holes
produced by the high energy scattering in the future collider was 
discussed in Ref.\cite{frol02-1,frol02-2,frol03-1}. Especially, in 
Ref.\cite{frol03-1} it was shown that superradiance for the bulk scalar
in the background of the $5d$ Myers-Perry rotating black hole\cite{myers86}
exists when the wave energy $\omega$ satisfies
$0 < \omega < m \Omega_a + k \Omega_b$, where $\Omega_a$ and 
$\Omega_b$ are angular frequencies of the black hole and, $m$ and $k$ are 
the azimuthal quantum numbers of the incident scalar wave. The generic 
conditions for the existence of the superradiance modes in the presence of 
single or multiple angular momentum parameters were derived 
recently\cite{jung05-2,jung05-3} when the incident bulk scalar, bulk 
electromagnetic and bulk gravitational waves are scattered by the higher-dimensional
rotating black hole.

Recently, the emission spectra for the brane fields were explored 
analytically\cite{ida03} in the low-energy regime and 
numerically\cite{harris05-1,ida05-1} in the entire range of the energy. 
The crucial difference of the brane fields from the bulk fields is the 
fact that the condition for the existence of the superradiance for the
brane fields is $0 < \omega < m \Omega$ while same condition for the 
bulk field is $0 < \omega < \sum_i m_i \Omega_i$ as shown in 
Ref.\cite{jung05-3}. Thus, in the background of the higher-dimensional black
holes carrying the multiple angular momentum parameters the bulk field can be
scattered superradiantly in the more wide range of $\omega$ compared to the
brane fields. This may change the standard claim that {\it black holes
radiate mainly on the brane}\cite{emp00}. The purpose of this paper is to 
explore this issue by choosing the $5d$ Myers-Perry rotating black hole 
with two angular momentum parameters $a$ and $b$ as a prototype.

In the following we will compute the absorption and emission spectra in the 
full range of $\omega$ for the brane scalar and bulk scalar. As a computational 
technique we will adopt an appropriate numerical technique which will be explained 
in detail later. Although the superradiant scattering takes place more 
readily for the bulk scalar, its energy amplification arising due to the superradiance 
is shown to be roughly $10^{-9} \%$ while that for the brane scalar is order of 
unity. This fact indicates that the effect of the superradiance does not
change the standard claim, {\it i.e.} {\it black holes radiate mainly
on the brane}. In order to compare the superradiant effects in $4d$ and 
$5d$, we carry out the calculation in the Appendix for the superradiant
scattering in the $4d$ Kerr background. The comparision reveals a big 
difference ($8$ orders of magnitude) for the energy amplification. 

The $5d$ rotating black hole derived by Myers and Perry is expressed by a
metric
\begin{eqnarray}
\label{met-bl}
ds_5^2&=& -dt^2 + \frac{r^2 \rho^2}{\bigtriangleup} dr^2 + \rho^2 d \theta^2
+ (r^2 + a^2) \sin^2 \theta d\phi^2 +
(r^2 + b^2) \cos^2 \theta d\psi^2 
                                    \\   \nonumber
& & \hspace{2.0cm}   + 
\frac{r_0^2}{\rho^2} (dt + a \sin^2 \theta d\phi + b \cos^2 \theta d\psi)^2
\end{eqnarray}
where $0 \leq \phi, \psi < 2 \pi$, $0 \leq \theta \leq \pi / 2$, 
\begin{eqnarray}
\label{bozo1}
\rho^2&=& r^2 + a^2 \cos^2 \theta + b^2 \sin^2 \theta
                                                  \\   \nonumber
\bigtriangleup&=& (r^2 + a^2) (r^2 + b^2) - r_0^2 r^2
\end{eqnarray}
and, $a$ and $b$ are two angular momentum parameters. The mass $M$, two angular
momenta $J_1$ and $J_2$ and the Hawking temperature $T_H$ are given by
\begin{equation}
\label{quants}
M = \frac{3 \pi r_0^2}{8} \hspace{1.0cm}
J_1 = \frac{2}{3} M a \hspace{1.0cm}
J_2 = \frac{2}{3} M b \hspace{1.0cm}
T_H = \frac{r_H^4 - a^2 b^2}{2\pi r_H (r_H^2 + a^2) (r_H^2 + b^2)}
\end{equation}
where $r_H$ is an horizon radius defined by $\bigtriangleup = 0$ at
$r = r_H$.

The induced $4d$ metric on the brane can be written as 
\begin{equation}
\label{met-br}
ds_4^2 = -dt^2 + \frac{r^2 \rho^2}{\bigtriangleup} dr^2 + \rho^2 d \theta^2
+ (r^2 + a^2) \sin^2 \theta d\phi^2
+ \frac{r_0^2}{\rho^2} (dt + a \sin^2 \theta d\phi)^2
\end{equation}
if the self-gravity on the brane is negligible, 
where we assume $0 \leq \theta \leq \pi$ to cover the whole $4d$ spacetime.
The scalar wave equation $\Box \Phi_{BR} = 0$ in the background 
(\ref{met-br}) is 
not separable. If, however, $b=0$, this wave equation is separable into the 
following radial and angular equations:
\begin{eqnarray}
\label{rad-ang-br} 
& & \frac{d}{d r} \left( \tilde{\bigtriangleup} \frac{d R_{BR}}{d r} \right)
+ \left[ \frac{[\omega (r^2 + a^2) - a m ]^2}{\tilde{\bigtriangleup}} - 
\Lambda_{\ell}^m \right] R_{BR} = 0
                                                  \\   \nonumber
& & \frac{1}{\sin \theta} \frac{d}{d \theta}
\left( \sin \theta \frac{d \Theta_{BR}}{d \theta} \right) + 
\left[ -\frac{m^2}{\sin^2 \theta} + \omega^2 a^2 \cos^2 \theta + 
{\cal E}_{\ell}^m \right] \Theta_{BR} = 0
\end{eqnarray}
where $\tilde{\bigtriangleup} = r^2 + a^2 - r_0^2$ and  
$\Lambda_{\ell}^m = {\cal E}_{\ell}^m + a^2 \omega^2 - 2 a m \omega$. When
deriving Eq.(\ref{rad-ang-br}), we used a factorization condition
$\Phi_{BR} = e^{-i \omega t} e^{-i m \phi} R_{BR}(r) \Theta_{BR} (\theta)$.  
 
The eigenvalue of the angular equation ${\cal E}_{\ell}^m$ was computed in
Ref.\cite{seid89} as an expansion of $a \omega$. Since, however, we need 
${\cal E}_{\ell}^m$ when $a \omega$ is arbitrarily large, we would like to 
solve the angular equation numerically. This is easily solved as following.
First, we note that $\Theta_{BR}$ becomes the usual spherical harmonics
$|\ell, m>$ when $a \omega = 0$. Of course, in this case 
${\cal E}_{\ell}^m = \ell (\ell + 1)$. When $a \omega \neq 0$, we expand
$\Theta_{BR}$ as $\Theta_{BR} = \sum_{\ell'} C_{\ell \ell'} |\ell', m>$. 
Then the 
angular equation reduces to the following eigenvalue equation
\begin{equation}
\label{angular1}
\sum_{\ell'} A_{\ell'' \ell'}^m C_{\ell \ell'} =
{\cal E}_{\ell}^m C_{\ell \ell''}
\end{equation}
where 
\begin{equation}
\label{angular2}
A_{\ell'' \ell'}^m = \ell' (\ell' + 1) \delta_{\ell' \ell''} - 
a^2 \omega^2 <\ell'', m|\cos^2 \theta|\ell', m>.
\end{equation}
Thus, the coefficients $C_{\ell \ell'}$ and the separation constant
${\cal E}_{\ell}^m$ are simultaneously obtained by computing the eigenvectors
and eigenvalues of the matrix $A_{\ell'' \ell'}^m$.
Solving the eigenvalue equation (\ref{angular1}) numerically, one can easily 
compute the $a \omega$-dependence of ${\cal E}_{\ell}^m$.

Now, we consider the wave equation $\Box \Phi_{BL} = 0$ for the bulk scalar
in the background of the $5d$ metric (\ref{met-bl}). The wave equation is always 
separable and the radial and angular equations are 
\begin{eqnarray}
\label{rad-ang-bl}
& & \hspace{5.0cm}
\frac{\bigtriangleup}{r} \frac{d}{d r}
\left( \frac{\bigtriangleup}{r} \frac{R_{BL}}{d r} \right) + W R_{BL} = 0
                                                   \\   \nonumber
& &\frac{d}{d \theta} \left( \sin \theta \cos \theta 
\frac{d \Theta_{BL}}{d\theta}
                  \right)
+ \left[\lambda_{\ell}^{m_1 m_2} - \omega^2 (a^2 \sin^2 \theta + b^2 \cos^2 
                                            \theta) 
- \frac{m_1^2}{\sin^2 \theta} - \frac{m_2^2}{\cos^2 \theta} \right]
\sin \theta \cos \theta \Theta_{BL} = 0
\end{eqnarray}
where
\begin{eqnarray}
\label{bl-poten}
& &W = \bigtriangleup \left[ -\lambda_{\ell}^{m_1 m_2} + \omega^2 
(r^2 + a^2 + b^2) + \frac{m_1^2 (a^2 - b^2)}{r^2 + a^2} + 
\frac{m_2^2 (b^2 - a^2)}{r^2 + b^2} \right]
                                             \\  \nonumber
& & \hspace{3.0cm}
+ r_0^2 (r^2 + a^2) (r^2 + b^2)
\left(\omega - \frac{m_1 a}{r^2 + a^2} - \frac{m_2 b}{r^2 + b^2} \right)^2.
\end{eqnarray}
When deriving Eq.(\ref{rad-ang-bl}), a factorization condition
$\Phi_{BL} = e^{-i \omega t} e^{-i (m_1 \phi + m_2 \psi}) R_{BL}(r) \Theta_{BL}
(\theta)$ is used. 

The angular equation can be solved numerically in a similar way to the case of 
the brane field. When $a = b = 0$, the eigenfunction of the angular equation is
expressed in terms of the Jacobi's polynomial as following\cite{frol03-1}
\begin{eqnarray}
\label{abgular3}
& &\Theta_{BL} \equiv |\ell, m_1, m_2> = 
2^{-(m_1 + m_2 - 1) / 2}
\sqrt{\frac{(2 \ell + m_1 + m_2 + 1) \Gamma [\ell + 1]
            \Gamma [\ell + m_1 + m_2 + 1]}
           {\Gamma [\ell + m_1 + 1] \Gamma [\ell + m_2 + 1]}}
                                                    \\   \nonumber
& & \hspace{5.0cm} \times
(1 - \cos 2\theta)^{m_1 / 2} (1 + \cos 2 \theta)^{m_2 / 2}
P_{\ell}^{(m_1,m_2)} (\cos 2\theta)
\end{eqnarray}
with $\lambda_{\ell}^{m_1 m_2} = (2\ell + m_1 + m_2) (2\ell + m_1 + m_2 + 2)$,
where $P_{\ell}^{(m_1,m_2)}$ is a jacobi's polynomial. When $a$ and $b$ are 
nonzero, we expand $\Theta_{BL}$ as 
$\Theta_{BL} = \sum_{\ell'} D_{\ell \ell'} |\ell', m_1, m_2>$. Then, by the 
same way as the brane case the angular equation reduces to the eigenvalue
problem:
\begin{equation}
\label{angular4}
\sum_{\ell'} B_{\ell'' \ell'}^{m_1 m_2} D_{\ell \ell'} =
\lambda_{\ell}^{m_1 m_2} D_{\ell \ell''}
\end{equation}
where
\begin{equation}
\label{abgular5}
B_{\ell'' \ell'}^{m_1 m_2} = (2 \ell' + m_1 + m_2) (2 \ell' + m_1 + m_2 + 2)
\delta_{\ell' \ell''} + 
<\ell'', m_1, m_2|\hat{H}_1|\ell', m_1, m_2>
\end{equation}
with $\hat{H}_1 = a^2 \omega^2 \sin^2 \theta + b^2 \omega^2 \cos^2 \theta$.
Solving the eigenvalue equation (\ref{angular4}) numerically, one can 
compute $\lambda_{\ell}^{m_1 m_2}$.

Now, we would like to discuss how to solve the radial equations in 
(\ref{rad-ang-br}) and (\ref{rad-ang-bl}). If we define $x = \omega r$ and 
$x_H = \omega r_H$, the radial equations reduce to
\begin{eqnarray}
\label{radial1}
& &(x^2 - x_H^2) \frac{d}{d x} (x^2 - x_H^2) \frac{d R_{BR}}{d x}
+ \left[ (x^2 + a^2 \omega^2 - a m \omega)^2 - \Lambda_{\ell}^m
         (x^2 - x_H^2) \right] R_{BR} = 0
                                                 \\  \nonumber
& & \hspace{3.0cm}
f(x, x_H) \frac{d}{d x} f(x, x_H) \frac{d R_{BL}}{d x} + \omega^4
   W R_{BL} = 0
\end{eqnarray}
where 
\begin{equation}
\label{radial2}
f(x, x_H) = \frac{\omega^4 \bigtriangleup}{x}
= \frac{x_H^2 (x^2 + a^2 \omega^2) (x^2 + b^2 \omega^2) - x^2
              (x_H^2 + a^2 \omega^2) (x_H^2 + b^2 \omega^2)}{x x_H^2}.
\end{equation}

The radial equations (\ref{radial1}) imply that if $R$ is a solution, 
$R^{\ast}$ is a solution too. The Wronskians between them become
\begin{eqnarray}
\label{wron1}
& &W[R_{BR}^{\ast}, R_{BR}]_x \equiv R_{BR}^{\ast} \frac{d R_{BR}}{d x}
- R_{BR} \frac{d R_{BR}^{\ast}}{d x} = \frac{{\cal C}_1}{x^2 - x_H^2}
                                                 \\   \nonumber
& &W[R_{BL}^{\ast}, R_{BL}]_x \equiv R_{BL}^{\ast} \frac{d R_{BL}}{d x}
- R_{BL} \frac{d R_{BL}^{\ast}}{d x} = 
\frac{{\cal C}_2 x}{(x^2 - x_H^2) (x^2 - a^2 b^2 \omega^4 / x_H^2)}
\end{eqnarray}
where ${\cal C}_1$ and ${\cal C}_2$ are integration constants.

From the radial equations (\ref{radial1}) one can derive the near-horizon
and asymptotic solutions analytically as a series form\cite{sanc78,jung04-2}.
The explicit expressions for the solutions of the radial equations 
convergent near horizon are
\begin{eqnarray}
\label{near-h1}
& & {\cal G}_{\ell,BR}^m (x, x_H) = e^{\rho_4 \ln |x - x_H|}
\sum_{n=0}^{\infty} d_{\ell,n}^m (x - x_H)^n
                                              \\   \nonumber
& & {\cal G}_{\ell,BL}^{(m_1,m_2)} (x, x_H) = e^{\rho_5 \ln |x - x_H|}
\sum_{n=0}^{\infty} d_{\ell, n}^{(m_1,m_2)} (x - x_H)^n
\end{eqnarray}
where
\begin{equation}
\label{near-h2}
\rho_4 = -i \frac{\omega (r_H^2 + a^2) (\omega - m \Omega_a)}{2 x_H}
\hspace{1.0cm}
\rho_5 = -i \frac{r_H (r_H^2 + a^2) (r_H^2 + b^2) (\omega - m_1 \Omega_a
- m_2 \Omega_b)}{2 (r_H^4 - a^2 b^2)}.
\end{equation}
In Eq.(\ref{near-h1}) we choosed the sign in the exponents so that
the solutions (\ref{near-h1}) become ingoing in the frame of reference 
of an observer co-rotating with a black hole.
In Eq.(\ref{near-h2}) $\Omega_a$ and $\Omega_b$ are the angular frequency of the
rotating black hole corresponding to the angular momentum parameters 
$a$ and $b$:
\begin{equation}
\label{freq-1}
\Omega_a = \frac{a}{r_H^2 + a^2}
\hspace{2.0cm}
\Omega_b = \frac{b}{r_H^2 + b^2}.
\end{equation}
The recursion relations for the coefficients $d_{\ell,n}^m$ and 
$d_{\ell, n}^{(m_1,m_2)}$ can be easily derived by inserting 
Eq.(\ref{near-h1}) into the radial equation (\ref{radial1}). Since the 
explicit expressions are too lengthy, we will not present them. 
It is important to note that when $\omega < m \Omega_a$, the imaginary part
of $\rho_4$ becomes positive. This implies that the near-horizon solution
for the brane wave equation becomes the outgoing wave with respect to 
an observer at infinity. This guarantees that
the superradiant scattering occurs at $0 < \omega < m \Omega_a$ for the 
brane field. As expected, the second equation in Eq.(\ref{near-h2})
implies that the superradiance exists for the bulk scalar at 
$0 < \omega < m_1 \Omega_a + m_2 \Omega_b$. Using Eq.(\ref{wron1}) one can 
show that the Wronskians between the near-horizon solutions are 
\begin{eqnarray}
\label{wron2}
& &W[{\cal G}_{\ell,BR}^{m \ast}, {\cal G}_{\ell,BR}^m]_x = -2 i \omega
\frac{(r_H^2 + a^2) (\omega - m \Omega_a)}{x^2 - x_H^2}
|g_{\ell}^m|^2
                                    \\    \nonumber
& &W[{\cal G}_{\ell,BL}^{(m_1,m_2) \ast}, {\cal G}_{\ell,BL}^{(m_1,m_2)}]_x =
-2 i \omega^2 \frac{(r_H^2 + a^2) (r_H^2 + b^2) (\omega - m_1 \Omega_a - m_2
\Omega_b) x}{r_H (x^2 - x_H^2) (x^2 - a^2 b^2 \omega^4 / x_H^2)}
|g_{\ell}^{(m_1,m_2)}|^2
\end{eqnarray}
where $g_{\ell}^m \equiv d_{\ell,0}^m$ and 
$g_{\ell}^{(m_1,m_2)} \equiv d_{\ell, 0}^{(m_1,m_2)}$. 

Next let us consider the solutions of the radial equations (\ref{radial1})
convergent at the asymptotic regime:
\begin{eqnarray}
\label{asymp1}
& &{\cal F}_{\ell (\pm), BR}^m (x, x_H) = (\pm i)^{\ell + 1} e^{\mp i x}
(x - x_H)^{\pm \rho_4} \sum_{n=0}^{\infty}
\tau_{n (\pm)}^{BR} x^{-(n+1)}
                                       \\   \nonumber
& &{\cal F}_{\ell (\pm), BL}^{(m_1,m_2)} (x, x_H) = (\pm i)^{\ell + 3/2}
\frac{e^{\mp i x} (x - x_H)^{\pm \rho_5}}{\sqrt{x}}
\sum_{n=0}^{\infty} \tau_{n (\pm)}^{BL} x^{-(n+1)}.
\end{eqnarray}
${\cal F}_{(+)}$ and ${\cal F}_{(-)}$ represent the ingoing and outgoing 
waves respectively. The recursion relations between the coefficients are 
not explicitly given here. With an aid of Eq.(\ref{wron1}) it is easy to show 
that the Wronskians between the asymptotic solutions are
\begin{eqnarray}
\label{wron3}
& &W[{\cal F}_{\ell (+), BR}^m, {\cal F}_{\ell (-), BR}^m]_x = 
\frac{2 i}{x^2 - x_H^2}
                                             \\   \nonumber
& &W[{\cal F}_{\ell (+), BL}^{(m_1,m_2)},{\cal F}_{\ell (-), BL}^{(m_1,m_2)}]_x
= \frac{2 i x}{(x^2 - x_H^2) (x^2 - a^2 b^2 \omega^4 / x_H^2)}.
\end{eqnarray}

Next, we would like to show how the coefficients $g_{\ell}^m$ and 
$g_{\ell}^{(m_1,m_2)}$ are related to the partial scattering amplitude. For
this we define the real scattering solutions $R_{\ell,BR}^m$ and 
$R_{\ell,BL}^{(m_1,m_2)}$, which behave as 
\begin{eqnarray}
\label{rsolution1}
& &R_{\ell,BR}^m \stackrel{x \rightarrow x_H}{\sim} g_{\ell}^m 
(x - x_H)^{\rho_4} [1 + O(x - x_H)]
                                           \\   \nonumber
& &R_{\ell,BL}^{(m_1,m_2)} \stackrel{x \rightarrow x_H}{\sim} 
g_{\ell}^{(m_1,m_2)} (x - x_H)^{\rho_5} [1 + O(x - x_H)]
                                            \\   \nonumber
& &R_{\ell,BR}^m \stackrel{x \rightarrow \infty}{\sim} i^{\ell + 1}
\frac{2 \ell + 1}{2 x}
\left[e^{-i x + \rho_4 \ln |x - x_H|} - (-1)^{\ell} S_{\ell}^m (x_H)
      e^{i x - \rho_4 \ln |x - x_H|} \right] + O(x^{-2})
                                             \\    \nonumber
& &R_{\ell,BL}^{(m_1,m_2)} \stackrel{x \rightarrow \infty}{\sim}
\sqrt{\frac{2}{\pi}} \frac{i^{\ell + 3 /2} (\ell + 1)^2}
                          {x^{3/2}}
\left[e^{-i x + \rho_5 \ln |x - x_H|} - (-1)^{\ell + 1/2}
S_{\ell}^{(m_1,m_2)} (x_H) e^{i x - \rho_5 \ln |x - x_H|} \right]
+ O(x^{-5/2})
\end{eqnarray}
where $S_{\ell}^m (x_H)$ and $S_{\ell}^{(m_1,m_2)} (x_H)$ are the scattering 
amplitudes for the brane and bulk scalars respectively. From the near-horizon behavior
we can understand the Wronskians between the real scattering solutions 
$W[R_{\ell,BR}^{m \ast}, R_{\ell,BR}^m]_x$ and 
$W[R_{\ell,BL}^{(m_1,m_2) \ast}, R_{\ell,BL}^{(m_1,m_2)}]_x$ are exactly
same with Eq.(\ref{wron2}) respectively. 

If we define the phase shifts 
$\delta_{\ell}^m (x_H) = (1 / 2i) \ln S_{\ell}^m (x_H)$ and
$\delta_{\ell}^{(m_1,m_2)} (x_H) = (1 / 2i) \ln S_{\ell}^{(m_1,m_2)} (x_H)$, 
the asymptotic behavior of $R_{\ell,BR}^m$ and 
$R_{\ell,BL}^{(m_1,m_2)}$ can be written as
\begin{eqnarray}
\label{rsolution2}
& &R_{\ell,BR}^m \stackrel{x \rightarrow \infty}{\sim} \frac{2 \ell + 1}{x}
e^{i \delta_{\ell}^m} \sin \left[x + i \rho_4 \ln |x - x_H| - \frac{\pi \ell}
                                                                   {2}  
+ \delta_{\ell}^m \right] + O(x^{-2})
                                        \\      \nonumber
& &R_{\ell,BL}^{(m_1,m_2)} \stackrel{x \rightarrow \infty}{\sim}
\sqrt{\frac{8}{\pi}} \frac{(\ell + 1)^2}{x^{3/2}}
e^{i \delta_{\ell}^{(m_1,m_2)}}
\sin \left[x + i \rho_5 \ln |x - x_H| - \pi \frac{\ell + 1/2}{2} + 
              \delta_{\ell}^{(m_1,m_2)} \right] + O(x^{-5/2}).
\end{eqnarray} 
Assuming that the phase shifts are the complex quantities, {\it i.e.}
$\delta_{\ell}^m \equiv \eta_{\ell}^m + i \beta_{\ell}^m$ and 
$\delta_{\ell}^{(m_1,m_2)} \equiv \eta_{\ell}^{(m_1,m_2)} + 
i \beta_{\ell}^{(m_1,m_2)}$, the Wronskians derived from the asymptotic
behavior (\ref{rsolution2}) are
\begin{eqnarray}
\label{wron4}
& &W[R_{\ell,BR}^{m \ast}, R_{\ell,BR}^m]_x = 
\frac{-i (2 \ell + 1)^2}{x^2 - x_H^2} e^{-2 \beta_{\ell}^m}
\sinh 2 \beta_{\ell}^m
                                       \\   \nonumber
& &W[R_{\ell,BL}^{(m_1,m_2) \ast}, R_{\ell,BL}^{(m_1,m_2)}]_x = 
\frac{-8 i (\ell + 1)^4 x}{\pi (x^2 - x_H^2) (x^2 - a^2 b^2 \omega^4 / x_H^2)}
e^{-2 \beta_{\ell}^{(m_1,m_2)}} \sinh 2 \beta_{\ell}^{(m_1,m_2)}.
\end{eqnarray}
Equating Eq.(\ref{wron4}) with Eq.(\ref{wron2}) yields
\begin{eqnarray}
\label{coeff1}
& &|g_{\ell}^m|^2 = \frac{\left(\ell + \frac{1}{2}\right)^2}
                         {\omega (r_H^2 + a^2) (\omega - m \Omega_a)}
                    (1 - e^{-4 \beta_{\ell}^m})
                                               \\   \nonumber
& &|g_{\ell}^{(m_1,m_2)}|^2 = 
\frac{2 (\ell + 1)^4 r_H}
     {\pi \omega^2 (r_H^2 + a^2) (r_H^2 + b^2) (\omega - m_1 \Omega_a - 
m_2 \Omega_b)}
(1 - e^{-4 \beta_{\ell}^{(m_1,m_2)}}).
\end{eqnarray}
In the first equation of Eq.(\ref{coeff1}) $|g_{\ell}^m|^2 > 0$ implies that
the greybody factor (or the transmission coefficient) 
$1 - e^{-4 \beta_{\ell}^m} \equiv 1 - |S_{\ell}^m|^2$ becomes negative when
$0 < \omega < m \Omega_a$, which is nothing but the superradiant scattering.
Similarly, the superradiance for the bulk scalar exists 
when $\omega$ satisfies $0 < \omega < m_1 \Omega_a + m_2 \Omega_b$, which is easily
deduced from the second equation of Eq.(\ref{coeff1}). 

Now, we would like to discuss how to compute 
the physical quantities such as absorption cross section and emission rate
from
$g_{\ell}^m$ and 
$g_{\ell}^{(m_1,m_2)}$.
For this discussion it is convenient to introduce new wave solutions 
$\tilde{R}_{\ell, BR}^{m}$ and $\tilde{R}_{\ell, BL}^{(m_1,m_2)}$, which
differ from $R_{\ell, BR}^{m}$ and $R_{\ell, BL}^{(m_1,m_2)}$ in their
normalization. They are normalized as
\begin{eqnarray}
\label{rsolution3}
& &\tilde{R}_{\ell, BR}^{m}(x) \stackrel{x \rightarrow x_H}{\sim}
(x - x_H)^{\rho_4} \left[1 + O(x - x_H)\right]
                                               \\   \nonumber
& &\tilde{R}_{\ell, BL}^{(m_1,m_2)}(x) \stackrel{x \rightarrow x_H}{\sim}
(x - x_H)^{\rho_5} \left[1 + O(x - x_H)\right].
\end{eqnarray}
Since ${\cal F}_{(+)}$ and ${\cal F}_{(-)}$ derived in Eq.(\ref{asymp1}) are
linearly independent solutions of the radial equations, we can gererally 
express these new wave solutions as a linear combination of 
${\cal F}_{(\pm)}$:
\begin{eqnarray}
\label{rsolution4}
& &\tilde{R}_{\ell, BR}^{m} = f_{\ell (-)}^m (x_H) {\cal F}_{\ell (+),BR}^m
(x, x_H) + f_{\ell (+)}^m (x_H) {\cal F}_{\ell (-),BR}^m (x, x_H)
                                               \\   \nonumber
& &\tilde{R}_{\ell, BL}^{(m_1,m_2)} = f_{\ell (-)}^{(m_1,m_2)} (x_H)
{\cal F}_{\ell (+),BL}^{(m_1,m_2)} (x, x_H) +
f_{\ell (+)}^{(m_1,m_2)} (x_H) 
{\cal F}_{\ell (-),BL}^{(m_1,m_2)} (x, x_H)
\end{eqnarray}
where the coefficients $f_{\pm}$ are called the jost functions.
Using Eq.(\ref{wron3}) one can compute the jost functions in the 
following:
\begin{eqnarray}
\label{jost1}
& &f_{\ell (\pm)}^m (x_H) = \pm \frac{x^2 - x_H^2}{2 i}
W[{\cal F}_{\ell (\pm),BR}^m, \tilde{R}_{\ell,BR}^m]_x
                                                     \\   \nonumber
& &f_{\ell (\pm)}^{(m_1,m_2)} (x_H) = \pm \frac{(x^2 - x_H^2) 
                                    (x^2 - a^2 b^2 \omega^4 / x_H^2)}{2 i x}
W[{\cal F}_{\ell (\pm),BL}^{(m_1,m_2)}, \tilde{R}_{\ell,BL}^{(m_1,m_2)}]_x.
\end{eqnarray}
Inserting the  explicit expressions of ${\cal F}_{(\pm)}$ into 
Eq.(\ref{rsolution4}) and comparing those with the asymptotic behavior of 
the real scattering solutions in Eq.(\ref{rsolution1}), one can derive the 
following relations:
\begin{eqnarray}
\label{jost2}
& &g_{\ell}^m (x_H) = \frac{\ell + \frac{1}{2}}{f_{\ell (-)}^m (x_H)},
\hspace{1.0cm}
S_{\ell}^m (x_H) = \frac{f_{\ell (+)}^m (x_H)}{f_{\ell (-)}^m (x_H)}
                                                      \\  \nonumber
& &g_{\ell}^{(m_1,m_2)} (x_H) = \sqrt{\frac{2}{\pi}}
\frac{(\ell + 1)^2}{f_{\ell (-)}^{(m_1,m_2)} (x_H)},
\hspace{1.0cm}
S_{\ell}^{(m_1,m_2)} (x_H) = 
\frac{f_{\ell (+)}^{(m_1,m_2)} (x_H)}{f_{\ell (-)}^{(m_1,m_2)} (x_H)}.
\end{eqnarray}

Combining Eq.(\ref{coeff1})  and Eq.(\ref{jost2}), we can compute the greybody
factors in terms of the jost functions:
\begin{eqnarray}
\label{grey-1}
& &{\cal T}_{\ell,BR}^m \equiv 1 - |S_{\ell}^m|^2 = 
\frac{\omega (r_H^2 + a^2) (\omega - m \Omega_a)}
     {|f_{\ell (-)}^m|^2}
                                       \\     \nonumber
& &{\cal T}_{\ell,BL}^{(m_1,m_2)} \equiv 1 - |S_{\ell}^{(m_1,m_2)}|^2 =
\frac{\omega^2 (r_H^2 + a^2) (r_H^2 + b^2) 
              (\omega - m_1 \Omega_a - m_2 \Omega_b)}
     {r_H |f_{\ell (-)}^{(m_1,m_2)}|^2}.
\end{eqnarray}
Thus if $0 < \omega < m \Omega_a$, ${\cal T}_{\ell,BR}^m$ becomes negative
which indicates the existence of the superradiance for the brane scalar.
Same is true when $0 < \omega < m_1 \Omega_a + m_2 \Omega_b$ for the 
bulk scalar.

The partial absorption cross section $\sigma_{\ell}^m$ for the brane scalar
and $\sigma_{\ell}^{(m_1,m_2)}$ for the bulk scalar are given by
\begin{eqnarray}
\label{pabs-1}
& &\sigma_{\ell}^m = \frac{\pi}{\omega^2} {\cal T}_{\ell,BR}^m
= \frac{\pi (r_H^2 + a^2) (\omega - m \Omega_a)}
       {\omega |f_{\ell (-)}^m|^2}
                                           \\   \nonumber
& &\sigma_{\ell}^{(m_1,m_2)} = \frac{4 \pi}{\omega^3}
                               {\cal T}_{\ell,BL}^{(m_1,m_2)}
= \frac{4 \pi (r_H^2 + a^2) (r_H^2 + b^2) 
              (\omega - m_1 \Omega_a - m_2 \Omega_b)}
       {x_H |f_{\ell (-)}^{(m_1,m_2)}|^2}.
\end{eqnarray}
Of course, the total absorption cross sections $\sigma_{BR}$ and 
$\sigma_{BL}$ are algebric sum of their partial absorption cross sections:
\begin{equation}
\label{tabs-1}
\sigma_{BR} = \sum_{\ell,m} \sigma_{\ell}^m
\hspace{2.0cm}
\sigma_{BL} = \sum_{\ell,m_1,m_2} \sigma_{\ell}^{(m_1,m_2)}.
\end{equation}

The total emission rate $\Gamma_{BR}$ for the brane scalar and $\Gamma_{BL}$
for the bulk scalar are given by
\begin{equation}
\label{temis-1}
\Gamma_{BR} = \sum_{\ell,m} \Gamma_{\ell}^m d \omega
\hspace{2.0cm}
\Gamma_{BL} = \sum_{\ell,m_1,m_2} \Gamma_{\ell}^{(m_1,m_2)} d \omega
\end{equation}
where
\begin{eqnarray}
\label{pemis-1}
& &\Gamma_{\ell}^m = \frac{1}{e^{(\omega - m \Omega_a) / T_H} - 1}
   \frac{\omega^3 \sigma_{\ell}^m}{2 \pi^2}
                                            \\   \nonumber
& &\Gamma_{\ell}^{(m_1,m_2)} = \frac{1}
                    {e^{(\omega - m_1 \Omega_a - m_2 \Omega_b) / T_H} - 1}
\frac{\omega^4 \sigma_{\ell}^{(m_1,m_2)}}{8 \pi^2}
\end{eqnarray}
and $T_H$ is an Hawking temperature given in Eq.(\ref{quants}). Therefore, 
we can compute all physical quantities related to the scattering between the
rotating black hole and the scalar field if we can compute the jost functions.

Now, we would like to present briefly how to compute the jost functions 
numerically. It is important to note that besides the near-horizon or 
asymptotic solution, we can derive the solutions from the radial equations 
(\ref{radial1}) which is convergent in the neighborhood of $x = b$, where
$b$ is an arbitrary point. Their expressions are 
\begin{eqnarray}
\label{inter-1}
& &\varphi_{\ell,BR}^m (x) = (x - x_H)^{\rho_4} 
\sum_{n=0}^{\infty} D_{\ell,n}^m (x - b)^n
                                         \\   \nonumber
& &\varphi_{\ell,BL}^{(m_1,m_2)} (x) = (x - x_H)^{\rho_5}
\sum_{n=0}^{\infty} D_{\ell,n}^{(m_1,m_2)} (x - b)^n.
\end{eqnarray}

\begin{figure}[ht!]
\begin{center}
\epsfysize=6.3 cm \epsfbox{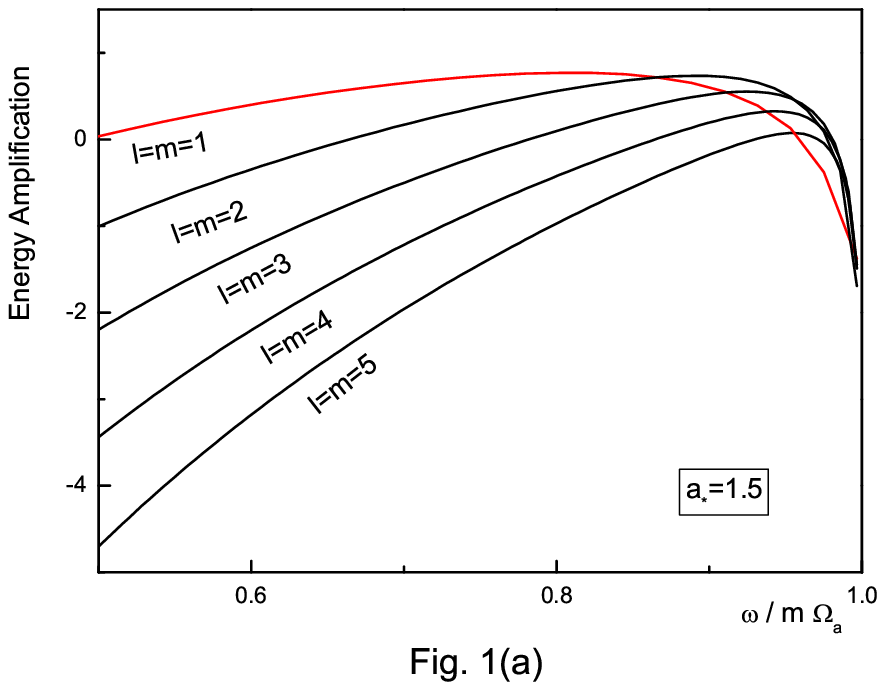}
\epsfysize=6.3 cm \epsfbox{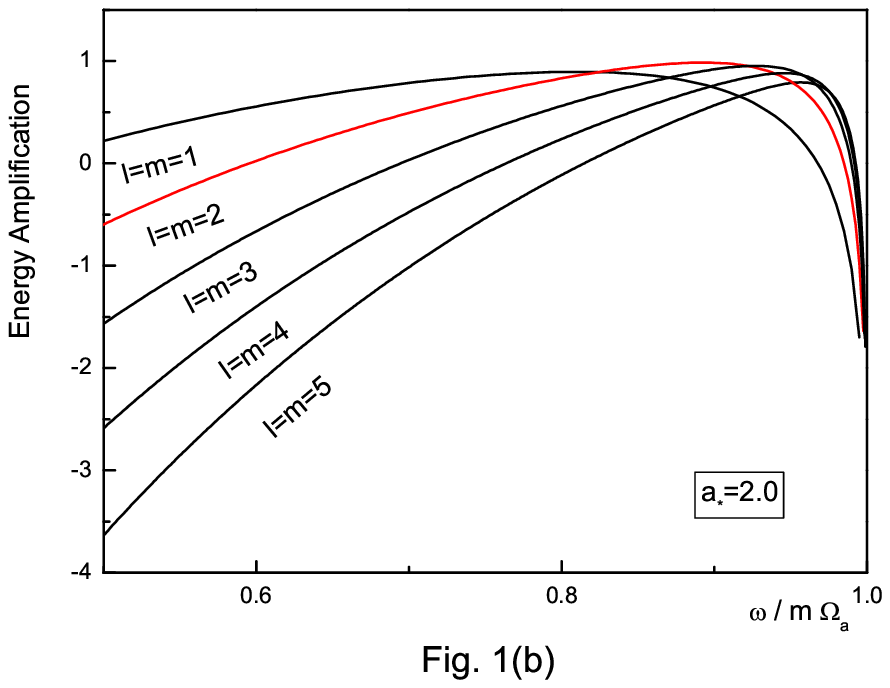}
\epsfysize=6.3 cm \epsfbox{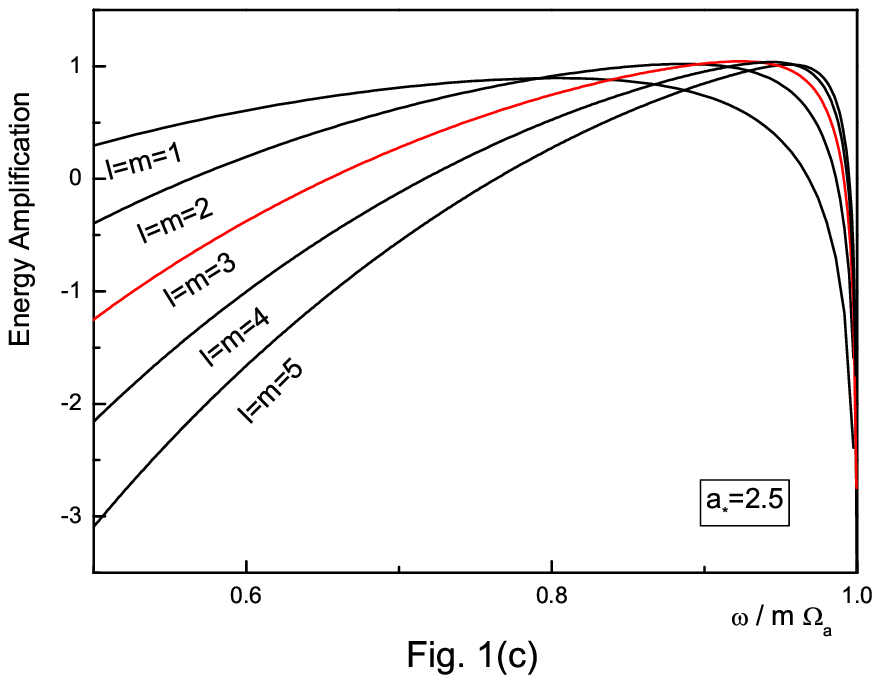}
\caption[fig1]{$Log$-plot of the energy amplification for the brane scalar when
$a_{\ast} \equiv a/r_H$ is $1.5$ (Fig. 1(a)), $2.0$ (Fig. 1(b)), and $2.5$ (Fig. 1(c)). When 
$a_{\ast} = 2.0$
(or $2.5$), the $\ell = m = 2$ (or $\ell = m = 3$) mode has a maximum peak. This means 
that the superradiant scattering of the higher modes becomes more and more significant
when $a_{\ast}$ becomes larger. This seems to be the important effect of the extra
dimensions. }                              
\end{center}
\end{figure}

\begin{figure}[ht!]
\begin{center}
\epsfysize=6.3 cm \epsfbox{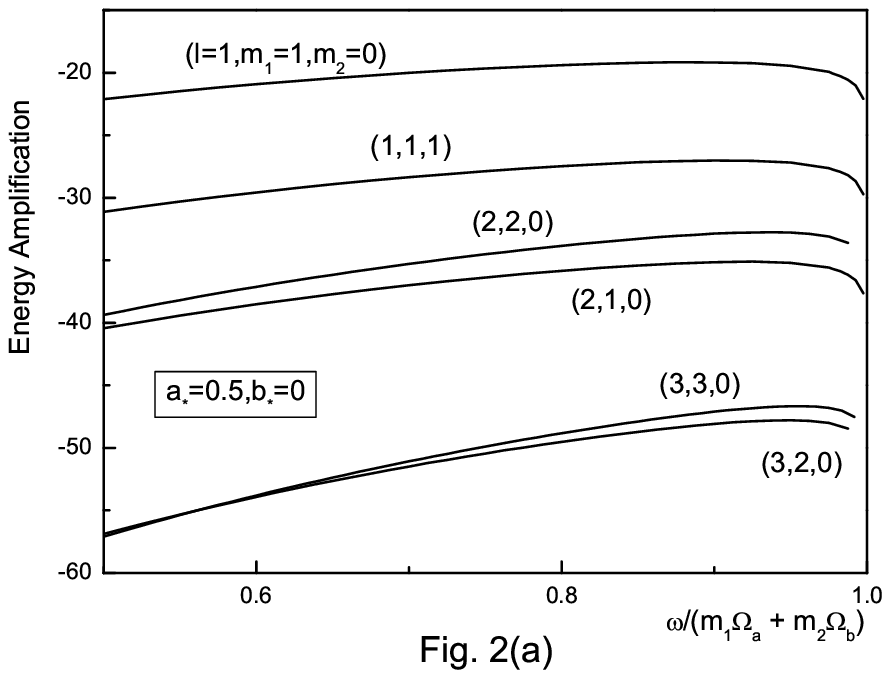}
\epsfysize=6.3 cm \epsfbox{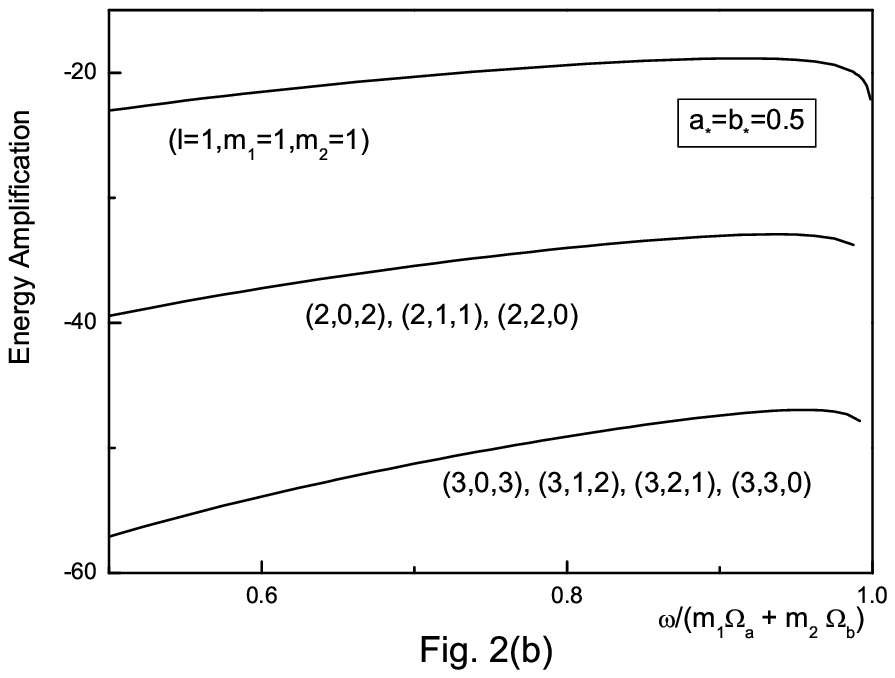}
\epsfysize=6.3 cm \epsfbox{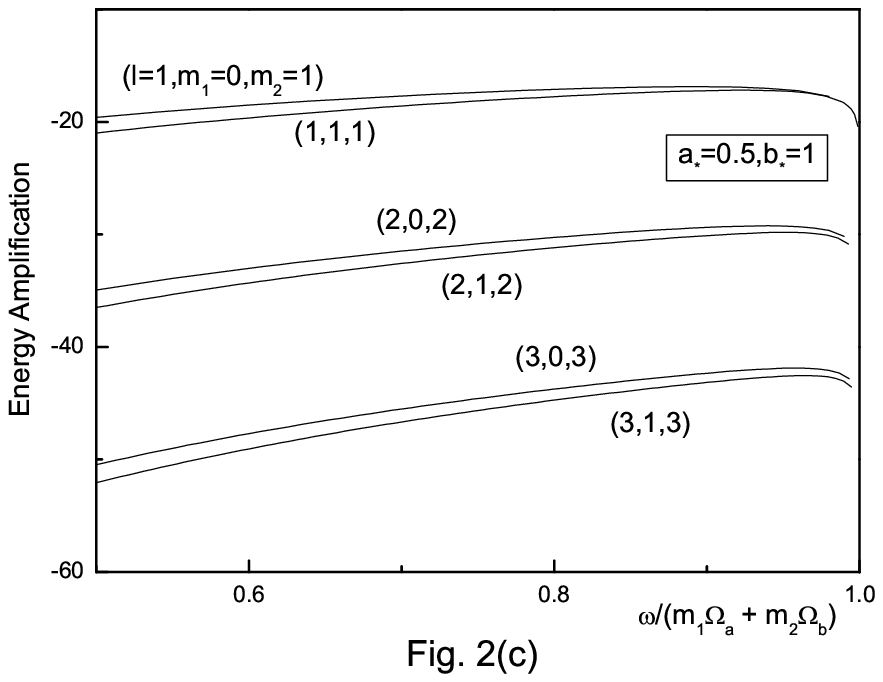}
\caption[fig2]{Energy amplification for the bulk scalar when 
$b_{\ast} \equiv b/r_H = 0$ (Fig. 2(a)),
$0.5$ (Fig. 2(b) and $1$ (Fig. 2(c)) with fixed $a_{\ast}$ as $0.5$. Usually the mode
which satisfies $m_1 + m_2 = \ell$ has a maximum amplification at fixed $\ell$. 
Comparision with Fig. 1 leads a conclusion that the effect of the superradiance for the 
bulk scalar is negligible.}
\end{center}
\end{figure}

\begin{figure}[ht!]
\begin{center}
\epsfysize=6.3 cm \epsfbox{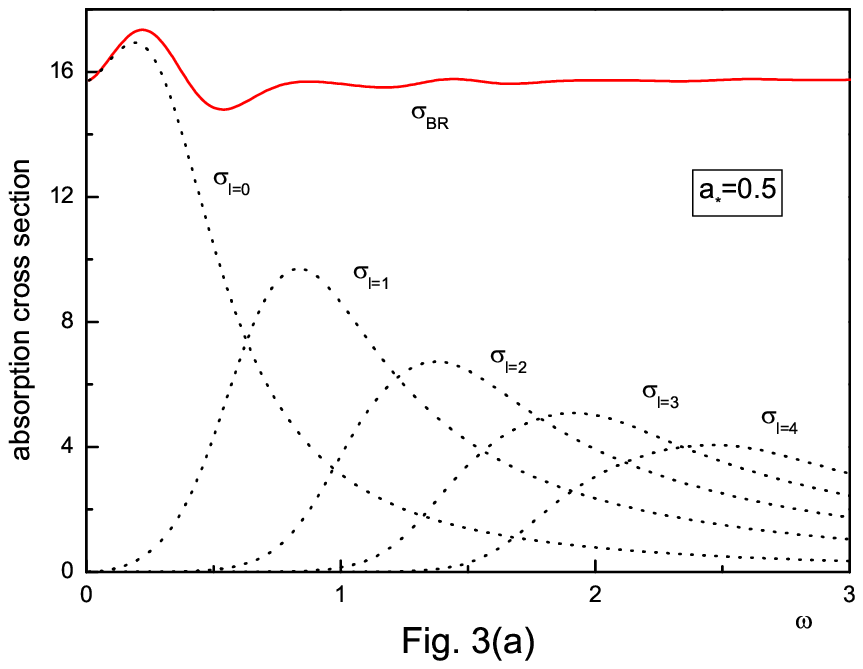}
\epsfysize=6.3 cm \epsfbox{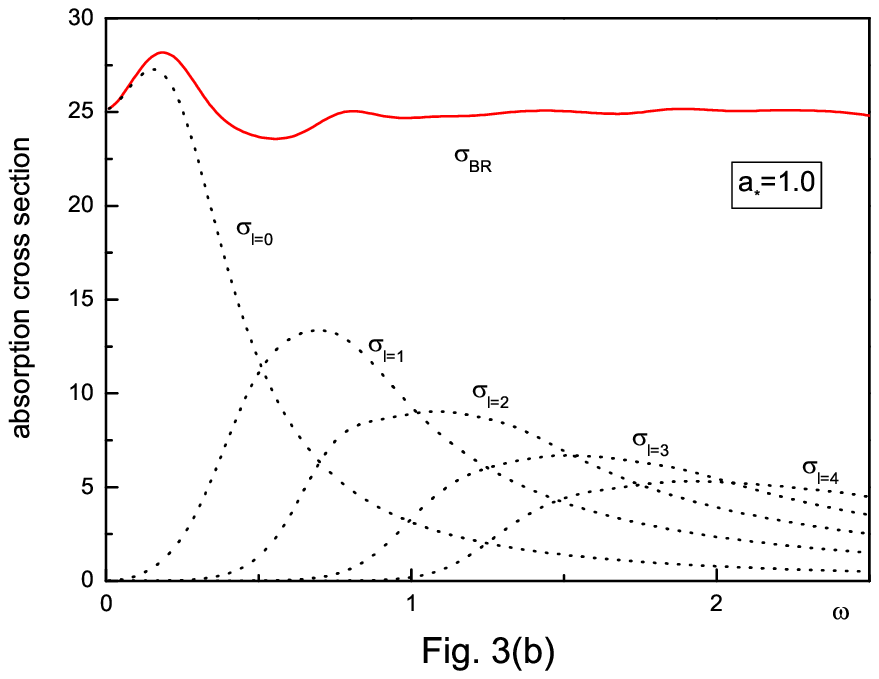}
\epsfysize=6.3 cm \epsfbox{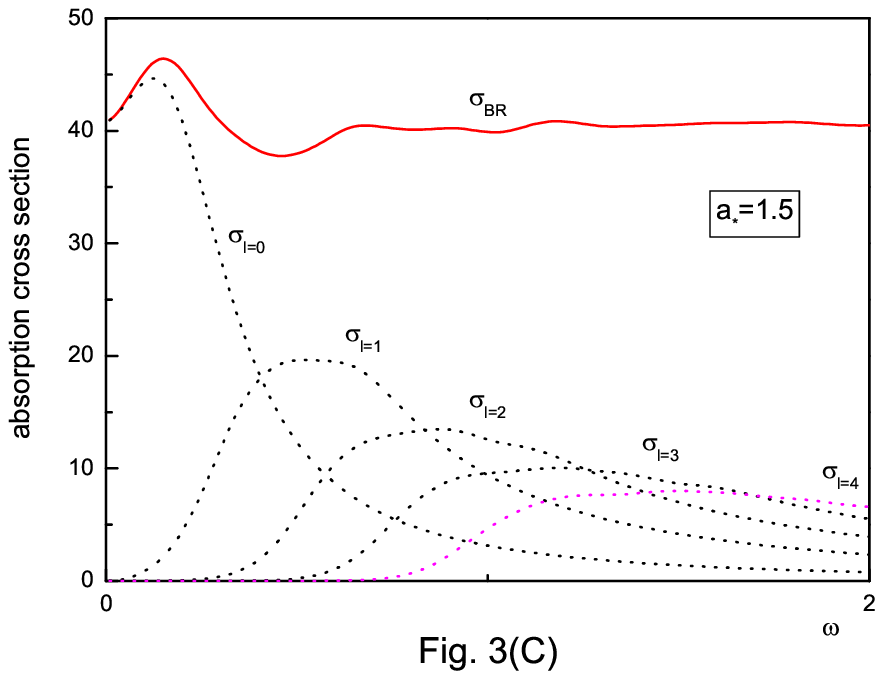}
\caption[fig3]{The total and partial absorption cross sections for the brane scalar
when $a_{\ast} = 0.5$ (Fig. 3(a), $1.0$ (Fig. 3(b)) and $1.5$ (Fig. 3(c)). Our 
numberical calculation shows that the low-energy limits of the total absorption cross 
sections always equal to the non-spherically symmetric horizon area 
$4\pi (r_H^2 + a^2)$. This fact gives rise to the conjecture that the universality 
for the scalar field discussed in Ref.\cite{das97} is extended to the non-spherically
symmetric spacetime. In the high-energy limits they approach to the nonzero values
which are roughly same with their low-energy limits.}
\end{center}
\end{figure}

The recursion relations between the coefficients can be explicitly derived 
by inserting Eq.(\ref{inter-1}) into the radial equation (\ref{radial1}),
which is not presented in this paper. Thus one can perform the matching
procedure between the near-horizon and the asymptotic solutions by making use
of this intermediate solutions as following. Matching procedure between the
near-horizon solutions and $\varphi_{\ell}$ generates a solution whose 
domain of convergence is larger than the near-horizon region. Repeat of 
this matching procedure would increase the convergence region more and more. 
Similar matching procedure between the asymptotic 
solutions and $\varphi_{\ell}$ can be repeated to decrease the convergent 
region from the asymptotic region. Eventually, we can obtain two solutions
which have common domain of convergence. Using these solutions we can
compute the jost functions with an aid of Eq.(\ref{jost1}).

Fig. 1 shows the $log$-plot of the superradiant scattering for the brane
field. The $y$-axis is a $\ln (-100 T_{\ell,BR}^m)$ for the most favour modes.
Fig. 1(a), (b), and (c) correspond to respectively $a_{\ast} = 1.5$, 
$a_{\ast} = 2.0$, and $a_{\ast} = 2.5$ where $a_{\ast} \equiv a / r_H$. As 
commented in Ref.\cite{harris05-1} the superradiant scattering for the 
higher modes becomes more and more significant as $a_{\ast}$ becomes
larger. This fact is evidently verified in Fig. 1. Table I shows the first 
three modes at each $a_{\ast}$ whose maximum energy amplification is large.
This Table shows that the lowest $\ell = m = 1$ mode has a 
maximum energy amplification at 
$a_{\ast} = 1.5$. But at $a_{\ast} = 2.0$ (or $2.5$) $\ell = m = 2$ (or 
$\ell = m = 3$) mode has a maximum amplification. It also shows that the average 
amplification tends to increase with increasing $a_{\ast}$. 
\begin{center}
{\large{Table I}}: Maximum Energy Amplification for the Several Modes of Brane Scalar 
\end{center}

\begin{center}
\begin{tabular}{|c|r|c|r|c|r|}
\hline
\multicolumn{2}{|c}  {$a_{\ast} = 1.5$} &
\multicolumn{2}{|c}  {$a_{\ast} = 2.0$} &    
\multicolumn{2}{|c|}  {$a_{\ast} = 2.5$} \\
\hline  \hline
modes & maximum energy & modes & maximum energy & modes  & maximum energy \\         & amplification (\%) &   & amplification (\%) &    & amplification (\%) \\
\hline
$(1,1)$  & $2.1604$ & $(2,2)$  & $2.6785$   & $(3,3)$ & $2.8399$    \\
$(2,2)$  & $2.0863$ & $(3,3)$  & $2.5934$   & $(4,4)$ & $2.8212$    \\
$(3,3)$  & $1.7392$ & $(1,1)$  & $2.4476$   & $(2,2)$ & $2.7748$    \\
\hline 
\end{tabular}
\\ \hspace{7.5cm} ($(p,q)$ means $\ell = p$ and $m = q$.)
\end{center}
\vspace{0.5cm}

Fig. 2 shows the $log$-plot of the superradiant scattering for the
bulk scalar. The vertical axis is a $\ln(-100 T_{\ell,BL}^{m_1,m_2})$
when $b_{\ast} = 0$, $0.5$ and $1$ with $a_{\ast} = 0.5$ where 
$b_{\ast} \equiv b/r_H$. The modes in this
figure are selected by comparing the maximum energy amplification at
fixed $\ell$. Usually one of the mode which satisfies $m_1 + m_2 = \ell$ has the 
largest maximum amplification at given $\ell$. When, especially, $a_{\ast} = b_{\ast}$,
the amplifications for the modes which have same $m_1 + m_2$ are 
exactly identical. For example, when $\ell = 2$, the amplications for 
$(0,2)$, $(1,1)$ and $(2,0)$ are exactly identical, where $(p,q)$ means
$m_1 = p$ and $m_2 = q$. When $a \neq b$, this kind of degeneracy is broken. 
However, still one of the modes which satisfies $m_1 + m_2 = \ell$ generally
has the 
largest maximum amplification. In Table II the maximum amplification values 
for several modes are given. Table I and Table II show that the energy 
amplification for the brane scalar is order of unity while that for the 
bulk scalar is order of $10^{-9} \%$
\footnote{The fact that the energy amplification for the bulk scalar is smaller
than for the brane scalar can be partly understood if we counter the power of 
the energy dependence. while the energy amplification for the brane scalar 
is proportional to $w^2$, that for the bulk scalar is proportional to $w^3$. 
Since the superradiant scattering usually takes place in the low-energy region,
we can conjecture that the energy amplification  for the bulk scalar can be
small. However, it does not explain the big difference between bulk and brane. 
It is unclear at least for us how to explain this issue physically.}. 
This means that the effect of the
superradiance for the bulk scalar is negligible. 
This indicates that consideration of the effect of the superradiance does not change 
the standard claim\cite{emp00} that 
{\it black holes radiate mainly on the brane}.
This will be confirmed in Fig. 5 explicitly.

\begin{center}
{\large{Table II}}: Maximum Energy Amplification for the Several Modes of Bulk Scalar
\end{center}

\begin{center}
\begin{tabular}{|r|r|r|r|r|r|}
\hline
\multicolumn{2}{|c} {$a_{\ast} = 0.5$, $b_{\ast} = 0$} &
\multicolumn{2}{|c}  {$a_{\ast} = 0.5$, $b_{\ast} = 0.5$} &
\multicolumn{2}{|c|}  {$a_{\ast} = 0.5$, $b_{\ast} = 1$} \\
\hline  \hline
modes & maximum energy & modes & maximum energy & modes  & maximum energy \\
     & amplification (\%) &   & amplification (\%) &    & amplification (\%) \\
\hline
$(1,1,0)$ & $4.723 \times 10^{-9}$  & $(1,1,1)$  & $6.494 \times 10^{-9}$   & $(1,0,1)$ & $4.748 \times 10^{-8}$    \\
$(1,1,1)$ & $1.849 \times 10^{-12}$  & $(1,0,1)$  & $4.682 \times 10^{-9}$   & $(1,1,1)$ & $3.456 \times 10^{-8}$    \\
$(2,2,0)$ & $5.933 \times 10^{-15}$  & $(2,0,2)$  & $5.035 \times 10^{-15}$   & $(2,0,2)$ & $2.013 \times 10^{-13}$    \\
$(2,1,0)$ & $5.632 \times 10^{-16}$  & $(2,1,2)$  & $4.961 \times 10^{-15}$   & $(2,1,2)$ & $1.121 \times 10^{-13}$    \\
$(3,3,0)$ & $5.444 \times 10^{-21}$  & $(3,0,3)$  & $4.019 \times 10^{-21}$   & $(3,0,3)$ & $6.492 \times 10^{-19}$    \\
$(3,2,0)$ & $1.768 \times 10^{-21}$  & $(3,1,3)$  & $3.540 \times 10^{-21}$   & $(3,1,3)$ & $3.323 \times 10^{-19}$    \\
\hline
\end{tabular}
\\ \hspace{5.5cm} ($(p,q,r)$ means $\ell = p$, $m_1 = q$ 
and $m_2 = r$.)
\end{center}
\vspace{0.5cm}

Fig. 3 shows the total and partial absorption cross sections for the brane scalar when
$a_{\ast} = 0.5$, $1$ and $1.5$. The partial absorption cross section plotted
in Fig. 3 is defined as $\sigma_{\ell} = \sum_m \sigma_{\ell}^m$. The negative
value of $\sigma_{\ell}^m$ for $m > 0$ in the range of 
$0 < \omega < m \Omega_a$ arising due to the superradiant scattering is 
compensated by the positive value of $\sigma_{\ell}^m$ for 
$m \leq 0$ in the same range arising due to the normal scattering. Therefore,
the partial absorption cross section $\sigma_{\ell}$ is positive in the full
range of $\omega$. 

\begin{figure}[ht!]
\begin{center}
\epsfysize=10.0 cm \epsfbox{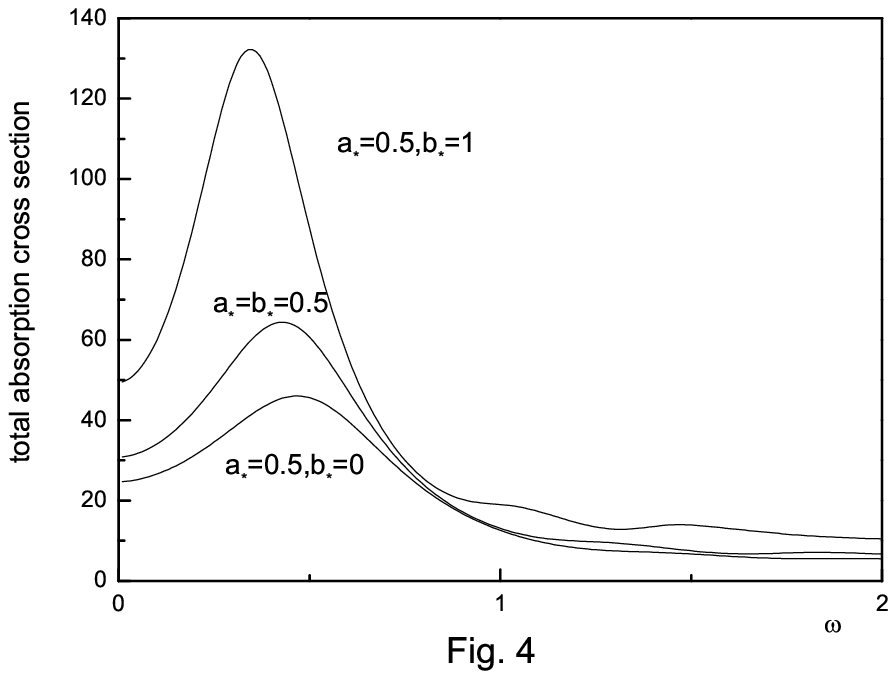}
\caption[fig4]{The total absorption cross sections for the bulk scalar when
$b_{\ast} = 0$, $0.5$, and $1.0$ with $a_{\ast} = 0.5$. The low-energy limits always
equal to the area of the non-spherically symmetric horizon hypersurface
$2\pi^2 (r_H^2 + a^2) (r_H^2 + b^2) / r_H$. Unlike the non-rotating black hole the 
contribution of the higher partial waves to the total absorption cross section is 
negligible.}
\end{center}
\end{figure}

\begin{figure}[ht!]
\begin{center}
\epsfysize=10.0 cm \epsfbox{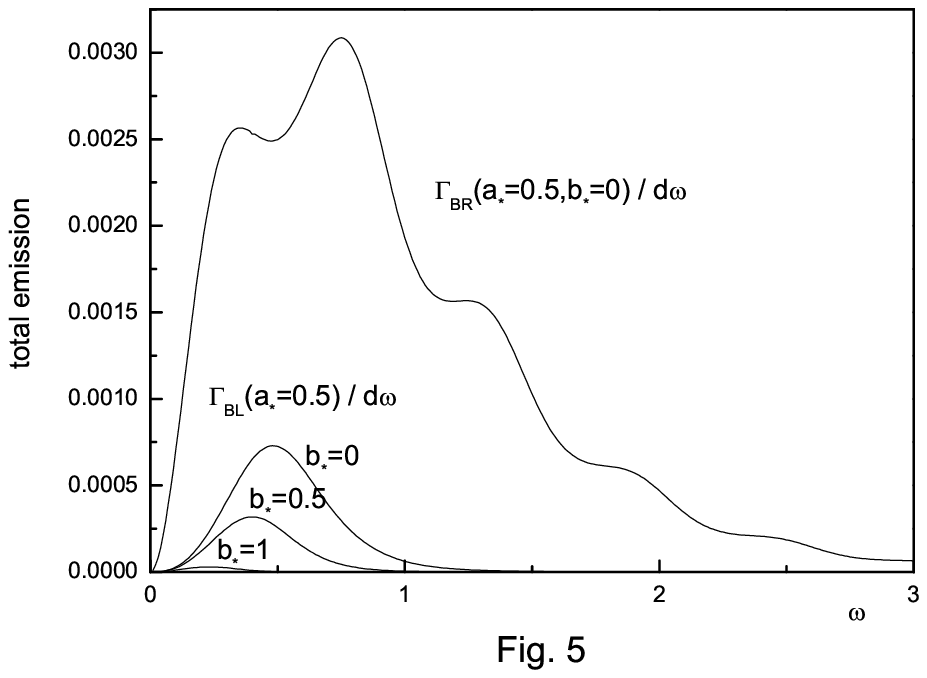}
\caption[fig5]{The total emission rate for the brane and bulk scalar fields. This figure 
shows that the $5d$ rotating black hole radiates mainly on the brane although we take 
the effect of the superradiance into account.}
\end{center}
\end{figure}

In the low-energy limits the total absorption cross sections exactly equal to 
the non-spherically symmetric horizon area ${\cal A}_{BR}$ defined
\begin{equation}
\label{br-horizon}
{\cal A}_{BR} \equiv \int_0^{\pi} d \theta \int_0^{2\pi} d\phi
\sqrt{g_{\theta \theta} g_{\phi \phi} - (g_{\theta \phi})^2} \Bigg|_{r = r_H}
= 4 \pi (r_H^2 + a^2).
\end{equation}
As proved in Ref.\cite{das97} the low-energy limit of the total absorption cross section
for the minimally coupled scalar always equals to the horizon area
in the asymptotically flat and spherically symmetric black hole, which is 
called `universality'. Although the general proof is not given yet, our 
numerical investigation supports the evidence that this universality seems 
to be extended to the non-spherically symmetric background. 

In the high-energy limits the total absorption cross sections approach to the 
nonzero values which are roughly same with the low-energy limits. In the 
intermediate region the total absorption cross sections do not exhibit an 
oscillatory pattern,
which seems to be the effect of the extra dimensions\cite{harris03-1,jung05-1}.

Fig. 4 shows the total absorption cross sections for the bulk scalar when 
$b_{\ast}=0$, $0.5$, and $1$ with $a_{\ast}=0.5$. The low-energy limits equal to the 
area of the non-spherically symmetric horizon hypersurface
${\cal A}_{BL}$ defined
\begin{equation}
\label{bl-horizon}
{\cal A}_{BL} \equiv \int_0^{\frac{\pi}{2}} d\theta
\int_0^{2\pi} d\phi d\psi 
\sqrt{g_{\theta \theta} \left[g_{\phi \phi} g_{\psi \psi}
          -\left(g_{\phi \psi}\right)^2 \right]}
= 2 \pi^2 \frac{(r_H^2 + a^2) (r_H^2 + b^2)}{r_H}.
\end{equation}
This result also supports that the universality in Ref.\cite{das97} holds in the
non-spherically symmetric background. In the high-energy limits the total 
absorption cross sections seem to approach to the nonzero values, which is 
much smaller than the low-energy limits. This fact indicates that unlike the
non-rotating black hole case\cite{harris03-1,jung05-1}, the contribution
of the higher partial waves except S-wave to the total absorption cross
section is too much negligible.

Fig. 5 shows the emission rate $\Gamma_{BL} / d\omega$ for the bulk scalar
and $\Gamma_{BR} / d\omega$ for the brane scalar together. For the brane
scalar we choosed $a=0.5$ and $b=0$ while for the bulk scalar $b$ is chosen
as $0$, $0.5$ and $1$ with $a = 0.5$. The wiggly pattern in 
$\Gamma_{BR} / d\omega$ indicates that unlike the non-rotating black holes 
the contribution of the higher
partial waves is not negligible. This means that the effect of the superradiant
scattering is crucially significant in the brane emission. This wiggly
pattern disappears in $\Gamma_{BL} / d\omega$, which means that the effect
of the superradiance is negligible. For the bulk scalar, therefore, the 
contribution of S-wave to the emission rate is dominant like the case of the 
non-rotating black hole background. Integrating the plots in Fig. 5, we can 
compute the total emission rate. For the brane scalar the total emission rate
is $0.00353832$ and for the bulk scalar $0.000343955$, $0.000123524$
and $7.44114 \times 10^{-6}$ for the cases of $b=0$, $b=0.5$ and $b=1$ 
respectively. Thus the emission rate for the bulk scalar is much
smaller than that for the brane scalar. Thus the effect of the superradiance in
$5d$ rotating black hole background does not seem to change the main conclusion 
of Ref.\cite{emp00}, {\it i.e.} {\it black holes radiate mainly on the brane}.

We computed the absorption and emission spectra for the brane and bulk scalar
fields when the 
spacetime is an $5d$ rotating black hole carrying the two different angular momentum
parameters. Although the effect of the superradiant scattering is taken into account,
the main conclusion of Ref.\cite{emp00} does not seem to be changed. This is due to the 
fact that the energy amplification for the bulk scalar is order of $10^{-9} \%$ while
that for the brane scalar is order of unity. It seems to be straightforward to extend
our calculation to the $6d$ rotating black hole background. It is of interest to check
explicitly whether or not the effect of the superradiance is negligible in $6d$ case. 

It is well-known that the Hawking radiation is highly dependent 
on the spin of the field. 
Thus,
it seems to be greatly important to take the effect of spin into 
account in the higher-dimensional rotating
black hole background. This is in progress and will be reported elsewhere.

\vspace{1cm}

{\bf Acknowledgement}:  
This work was supported by the Korea Research
Foundation under Grant (KRF-2003-015-C00109).

\newpage
\begin{appendix}{\centerline{\bf Appendix}}
\setcounter{equation}{0}
\renewcommand{\theequation}{A.\arabic{equation}}
The $4d$ Kerr metric is well-known in the form
\begin{eqnarray}
\label{app1}
ds^2 = &-& (1 - \frac{\mu r}{\Sigma}) dt^2 
        - \frac{2 a \mu r \sin^2 \theta}{\Sigma} dt d \phi
        + \frac{\Sigma}{\Delta} dr^2 + \Sigma d \theta^2  \\  \nonumber
 &+& (r^2+a^2 + \frac{a^2 \mu r \sin^2 \theta}{\Sigma}) \sin^2 \theta d \phi^2
\end{eqnarray}
where $\Delta = r^2 - \mu r + a^2 \equiv (r - r_+)(r - r_-)$,
$\Sigma = r^2 + a^2 \cos^2 \theta$ and
$r_{\pm} = (\mu \pm \sqrt{\mu^2 - 4 a^2})/2$ are the inner and outer horizons.
Then it is not difficult to show that the scalar wave equation 
$\Box \Phi = 0$ in this background is separable. The radial and angular
equations of this wave equation reduce to
\begin{eqnarray}
\label{app2}
&& \Delta \frac{d}{dr} \Delta \frac{dR}{dr} + [(r^2 w + a^2 w - am)^2
      - \Delta \Lambda_{\ell}^m ] R = 0     \\ \nonumber
&& \frac{1}{\sin \theta} \frac{d}{d \theta} (\sin \theta \frac{dT}{d \theta})
    +[- \frac{m^2}{\sin^2 \theta} + a^2 w^2 \cos^2 \theta + E_{\ell}^m ]T =0
\end{eqnarray}
where $\Phi = R(r) T(\theta) e^{i m \phi} e^{- i w t}$, 
$\Lambda_{\ell}^m = E_{\ell}^m + a^2 w^2 - 2 a m w$ and $E_{\ell}^m $ is
a separation constant.
Defining $x = w r$ and $x_{\pm} = w r_{\pm}$, one can show that
the radial equation becomes
\begin{eqnarray}
\label{app3}
(x - x_+)(x &-& x_-) \frac{d}{dx} (x - x_+)(x- x_-) \frac{dR}{dx} \\ \nonumber
&+& [(x^2 + a^2 w^2 - a m w )^2 - \Lambda_{\ell}^m (x - x_+)(x - x_-)]R = 0.
\end{eqnarray}
Solving the radial equation as a series form, one can derive the
near-horizon solution
\begin{equation}
\label{app4}
{\cal G}_{\ell}^m (x, x_+, x_-) = e^{\lambda_4 ln | x - x_+|} 
\sum_{n=0}^{\infty} d_{\ell, n}^m (x - x_+)^n
\end{equation}
and the asymptotic solution
\begin{equation}
\label{app5}
{\cal F}_{\ell (\pm)}^m (x, x_+, x_-) = (\pm i)^{\ell + 1} e^{\mp i x}
    (x - x_+)^{\pm \lambda_4} \sum_{n=0}^{\infty} \tau_{n (\pm)} x^{-(n+1)}
\end{equation}
where the recursion relations for $d_{\ell, n}^m$ and $\tau_{n (\pm)}$
can be explicitly derived by inserting (\ref{app4})  and (\ref{app5}) 
into (\ref{app3}). The factor $\lambda_4$ arises due
to the regular singular nature of the radial equation and its explicit 
expression is
\begin{equation}
\label{app6}
\lambda_4 = - i \frac{w (r_+^2 + a^2)(w - m \Omega)}{x_+ - x_-} 
\end{equation}
where $\Omega = a/(r_+^2 + a^2)$ is an angular frequency of the black hole.
If $0 < w < m \Omega$, $Im \lambda_4$ becomes negative which indicates that 
the near-horizon solution (\ref{app4}) becomes outgoing wave. Thus the 
superradiant scattering takes place under the condition  
$0 < w < m \Omega$.

\begin{figure}[ht!]
\begin{center}
\epsfysize=10.0 cm \epsfbox{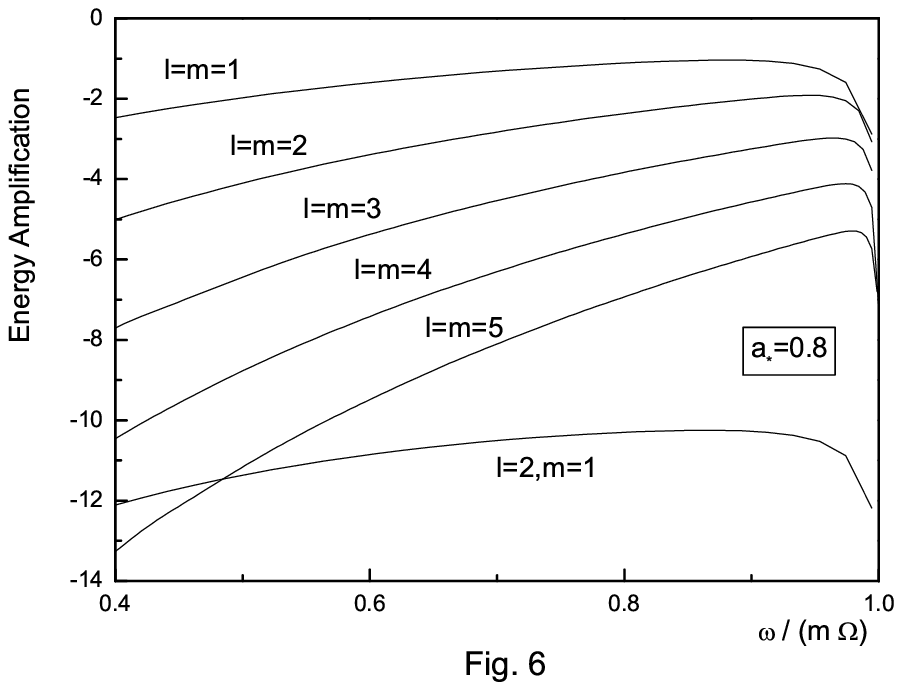}
\caption[fig6]{$Log$-plot of the energy amplification for the minimally-coupled
scalar when the spacetime
background is a $4d$ Kerr black hole with $a_{\ast} \equiv a/r_+ = 0.8$. 
This figure indicates that the
energy amplification for the $4d$ case is order of $10^{-1}\%$
like the maximally rotating case.}
\end{center}
\end{figure}

The energy amplification arising due to the superradiant scattering was
computed in Ref. \cite{press72,harris05-1} when the spacetime 
background is a maximally rotating ($a_{\ast} \equiv a/r_+ = 1$) 
Kerr black hole. The authors in Ref. \cite{press72,harris05-1} 
solved the radial and angular equations 
(\ref{app2}) directly by adopting the different numerical technique. 
For our case, however, the near-horizon and asymptotic solutions (\ref{app4}) 
and (\ref{app5}) are used.
Thus our numerical method cannot be applied to the case of the maximally 
rotating black hole because 
$a_{\ast} = 1$ implies the extremal limit, {\it i.e.} $r_+ = r_-$ and
$\lambda_4$ goes to infinity in this limit.
Applying the numerical method used in the present paper 
the energy amplification 
can be straightforwardly computed for $a_{\ast} < 1$.

Fig. 6 is a $\log$-plot of the energy amplification when $a_{\ast} = 0.8$.
Although 
Fig. 6 is different from Fig. 1 of Ref. \cite{press72} due to the
different choice of $a_{\ast}$, it indicates that the energy amplification 
of the scalar wave in the $4d$ Kerr black hole is order of $10^{-1} \%$ like 
the case of 
$a_{\ast} = 1$.

\end{appendix}

\end{document}